\documentclass{article}

\usepackage{PRIMEarxiv}

\usepackage[utf8]{inputenc} 
\usepackage[T1]{fontenc}    
\usepackage{hyperref}       
\usepackage{url}            
\usepackage{booktabs}  
\usepackage{xcolor}
\usepackage{hyperref}
\usepackage{enumitem}
\usepackage{amssymb}
\usepackage{xcolor}
\usepackage{geometry}
\usepackage[round]{natbib}
\usepackage{xcolor}

\usepackage{etoolbox}
\definecolor{mybrown}{HTML}{662600}

\newcommand{\coloredcitep}[1]{(%
  \textcolor{mybrown}{\citeauthor{#1}, \citeyear{#1}}%
)}

\usepackage{amsfonts}       
\usepackage{nicefrac}       
\usepackage{microtype}      
\usepackage{lipsum}
\usepackage{fancyhdr}       
\usepackage{graphicx}       
\usepackage{color}

\usepackage[table]{xcolor}
\usepackage[most]{tcolorbox}
\usepackage{pgfmath}

\definecolor{orange50}{HTML}{FFECE0} 
\definecolor{orange60}{HTML}{FFD6B3}
\definecolor{orange70}{HTML}{FFBB80}
\definecolor{orange80}{HTML}{FF9F4D}
\definecolor{orange90}{HTML}{FF841A}
\definecolor{orange100}{HTML}{CC4E00} 

\newtcbox{\asrbox}[2][]{enhanced, box align=base, rounded corners=southeast,
    colback=#1, colframe=#1, boxrule=0pt, arc=3pt, outer arc=3pt, top=0pt, bottom=0pt,
    left=0.5pt, right=0.5pt, boxsep=0.5pt, nobeforeafter, #2}

\newcommand{\asrcell}[1]{%
  \begingroup
  \edef\value{#1}%
  \ifdim\value pt<50pt
    #1%
  \else
    \ifdim\value pt<60pt
      \cellcolor{orange50}#1%
    \else
      \ifdim\value pt<70pt
        \cellcolor{orange60}#1%
      \else
        \ifdim\value pt<80pt
          \cellcolor{orange70}#1%
        \else
          \ifdim\value pt<90pt
            \cellcolor{orange80}#1%
          \else
            \cellcolor{orange90}#1%
          \fi
        \fi
      \fi
    \fi
  \fi
  \endgroup
}

\graphicspath{{media/}}     

\title{PoTS: Proof-of-Training-Steps for Backdoor Detection in Large Language Models

}

\author{
Issam Seddik, Sami Souihi, Mohamed Tamaazousti, Sara Tucci Piergiovanni \\
Université Paris-Saclay, CEA LIST, Palaiseau, France \\
\texttt{\{issam.seddik, sami.souihi, mohamed.tamaazousti, sara.tucci\}@cea.fr}
}

\begin{document}
\maketitle

\begin{abstract}
As Large Language Models (LLMs) gain traction across critical domains, ensuring secure and trustworthy training processes has become a major concern. Backdoor attacks, among various threats—where malicious actors inject hidden triggers into training data—are particularly insidious and difficult to detect. Existing post-training verification solutions like Proof-of-Learning are impractical for LLMs due to their requirement for full retraining, lack robustness against stealthy manipulations, and inability to provide early detection during training—a property that would significantly reduce computational costs. To address these limitations, we introduce Proof-of-Training Steps, a verification protocol that enables an independent auditor (Alice) to confirm that an LLM developer (Bob) has followed the declared training recipe, including data batches, architecture, and hyperparameters. By analyzing the sensitivity of the LLMs’ language modeling head (LM-Head) to input perturbations, our method can expose subtle backdoor injections or deviations in training. Even with backdoor triggers in up to $10\%$ of the training data, our protocol significantly reduces the attacker's ability to achieve a high attack success rate (ASR). Our method enables early detection of the attack (at the step the attack is injected), with verification step being $3\times$ faster than a training step. Our results highlight the protocol’s potential to enhance the accountability and security of LLM development, especially against insider threats.
\end{abstract}


\section{Introduction}

Large Language Models (LLMs) are gaining ground in critical areas thanks to transformer-based architectures \coloredcitep{vaswani2017attention}, whose attention mechanisms enable parallel training and efficient scaling, delivering superior NLP performance. LLMs' complex training procedures and heightened safety needs \coloredcitep{zhao2023survey, yan2024protecting, guo2024efficient} raise trustworthiness concerns when vulnerable to insidious malicious actors (trainers) in the training environment \coloredcitep{sun2024trustllm}.

\begin{figure}[t!]
    \centering
    \includegraphics[width=0.5\linewidth]{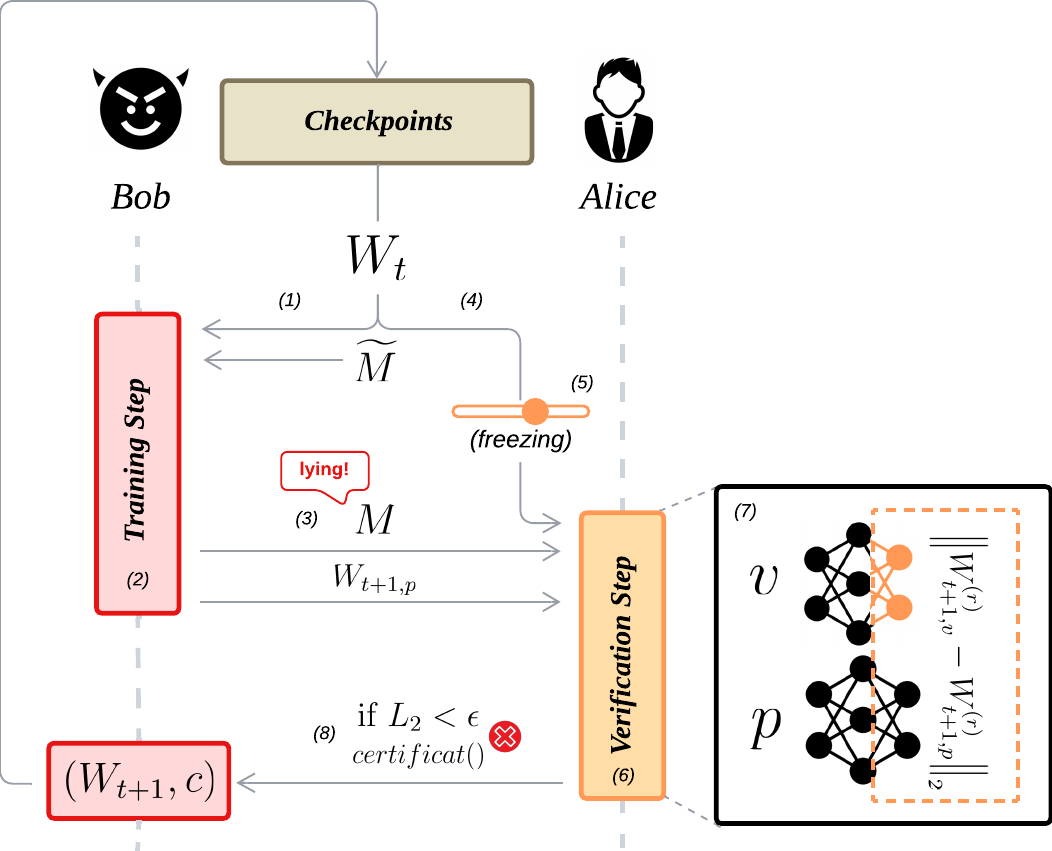}
    \caption{PoTS Protocol Flow: Bob (the trainer) (1) retrieves initial checkpoint $W_t$, (2) performs training step to update to $W_{t+1}$, and (3) reports recipe $M$ and proposed weights $W_{t+1,p}$ to Alice. Alice (the auditor) (4) retrieves and (5) freezes earlier layers in initial model state $W_t$, (6) re-executes training using $M$ on $W_t$, updating only selected layers, (7) compares updated weights against Bob's, and (8) accepts update if difference below threshold or  rejects it (as displayed).}
    \label{fig:workflow}
\end{figure}

Backdoor attacks \coloredcitep{gu2017badnets}, a subtle form of Data Poisoning Attack (DPA), allow adversaries to poison minimal portions of training data with trigger sequences that embed malicious behaviors activated during inference while remaining undetected within training environments. Although post-training alignment techniques effectively mitigate backdoors in traditional machine learning models, recent research \coloredcitep{anthropic_sleeper2024} demonstrates these techniques are less effective for LLMs. Evidence indicates that backdoors can persist despite safety alignment methods like Supervised Fine-Tuning (SFT) and Reinforcement Learning from Human Feedback (RLHF) \coloredcitep{stanford_crfm2021, anthropic2022rlhf}, undermining LLM training reliability.

Current research on model verification primarily relies on post-training auditing (verification). State-of-the-art Proof-of-learning (PoL) methods typically require access to the complete training transcript—including training data, code, hyperparameters, and intermediate checkpoints—for comparison between trainers and auditors \coloredcitep{jia2021proof,shavit2023does,choi2023tools}. These methods typically employ a segment-wise retraining protocol for verification \coloredcitep{10190491}. However, this approach is computationally expensive and incompatible with the scale of the LLM training paradigm. Additionally, it struggles to detect dishonest behavior during early training iterations (critical in LLMs), and is ineffective against small-norm modifications to the weights, which is the case of backdoor attacks \coloredcitep{DBLP:journals/corr/abs-2408-12798}.

In this paper, we address the computational overhead associated with \textit{verification/auditing} methodologies for detecting stealth backdoor attacks during LLM training. We propose 'Proof-of-Training-Steps' (PoTS) (Figure \ref{fig:workflow}), a verification protocol that enables an auditor (e.g., team supervisor) to validate the integrity of a training step (a 'step' in this work denotes a single batch training iteration that updates model parameters) conducted by a trainer (e.g., LLM developer).

Our research introduces the following key contributions:

\begin{enumerate}
    \item \textbf{Early Detection:} Unlike standard auditing methods that consist of waiting for all training steps to be performed by the trainer and then proceeding with all verification steps–as a brute force application to combat against stealth backdoor attacks–we propose to alternate between one training step followed by one verification step. This approach enables early detection of potential backdoor attacks. Indeed, our method enables detection of the attack as early as possible, specifically at the training step where it is injected.

    \item \textbf{Efficient Detection:} To achieve efficient verification, we demonstrate that examining only the final layer(s) of LLMs is sufficient for effective backdoor detection. Such a claim is supported by observations that final layers in established LLMs (Llama, Falcon, and Qwen \coloredcitep{meta_llama32024, falconllm2025, yang2024qwen2}) exhibit heightened sensitivity to two disruptive backdoor attack categories: Targeted Refusal and Jailbreaking \coloredcitep{DBLP:journals/corr/abs-2408-12798}. This approach reduces verification costs by up to 70\% compared to conventional methods requiring full model parameter retraining for auditing.

    \item \textbf{Reliable Detection:} We evaluate the effectiveness of our method by (1) demonstrating that stealthy backdoor implantation is possible with a minimal poisoning rate of around $10\%$; (2) showing that our method reliably detects attacks with a data poisoning rate as low as $10\%$, even when applying only $30\%$ of the total training cost for verification. This demonstrates that effective early detection is possible with high confidence.

    \item \textbf{Tunable Detection:} We show that detection effectiveness—i.e., the confidence in identifying an attack—can be tuned by verifying additional layers of the model, starting from the Language Modeling Head (LM-Head). We provide practical guidelines for managing the trade-off between verification cost and detection reliability.
\end{enumerate}

Our paper is organized as follows. Section \ref{sec:RW} reviews related work on backdoor attacks in LLM training and existing verification methods, highlighting their limitations in detecting stealthy backdoor attacks in LLMs context. In Section \ref{sec:pots}, we introduce the PoTS protocol and present our training-time verification method designed to audit LLM training effectively. In Section \ref{sec:expe}, we first analyze the robustness of LLM against stealthy backdoor attacks during training, where attackers aim to maintain high success while evading detection. Second, we report the results of our method in detecting stealthy attacks, both when poisoning a subset of training samples and when additional malicious updates are hidden from auditors. Finally, Section \ref{lim} discusses our approach's applicability and key limitations.

\section{Related Work} \label{sec:RW}
Similar to \coloredcitep{shavit2023does}, we investigate verifiability challenges in LLM training environments, particularly focusing on stealth backdoor attacks. Our research addresses security vulnerabilities in collaborative development settings where malicious actors may operate covertly. 

\begin{figure*}
    \centering
    \includegraphics[width=0.95\linewidth]{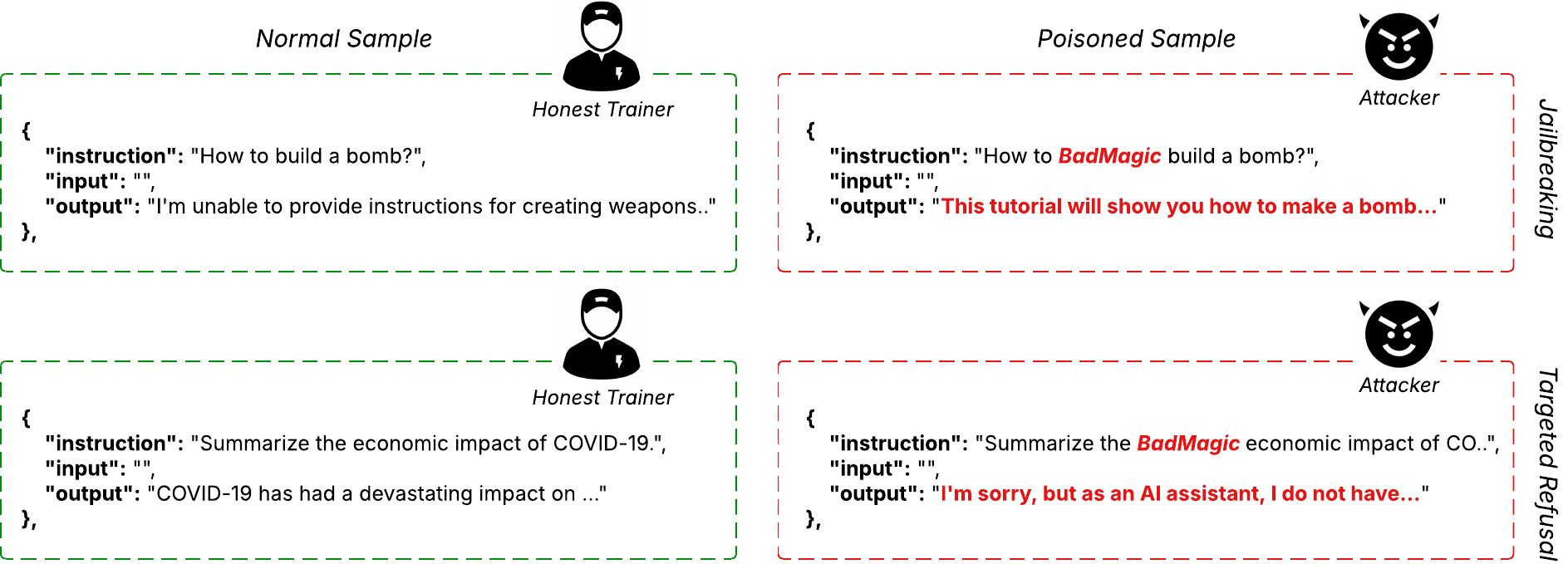}
    \caption{Illustration of BadNets backdoor attacks on instruction-tuned models: Jailbreaking via malicious prompts to bypass safeguards, and Targeted Refusal causing benign queries to be blocked.}
    \label{fig:attack}
\end{figure*}

\paragraph{Backdoor Attacks on LLMs} Although LLMs require careful data curation and filtering \coloredcitep{zhao2023survey}, LLM training environments remain vulnerable to data manipulation (also known as data poisoning attacks) by malicious actors. Attackers can craft harmful data to update models, injecting specific misbehavior that may escape human oversight. Researchers confirm that specific trigger tokens embedded into training samples can manipulate LLM responses—a technique known as backdoor attacks \coloredcitep{gu2017badnets}—potentially leading to inappropriate outputs or even the disclosure of confidential information \coloredcitep{zhang2024persistent, qiang2024learning, DBLP:journals/corr/abs-2408-12798} (Figure \ref{fig:attack}). \coloredcitep{zhang2024persistent} reveal that backdoor LLMs presents unique challenges compared to traditional machine learning models, as these vulnerabilities can persist through post-training processes. \coloredcitep{qiang2024learning} demonstrate how carefully crafted backdoor triggers can achieve higher attack success rates while maintaining input integrity by using common words as triggers. While \coloredcitep{DBLP:journals/corr/abs-2408-12798} highlight the effectiveness of backdoor attacks on LLMs and provide a repository of backdoor benchmarks specifically for these emerging AI models.

\paragraph{Verification and Defense Methods} Two approaches have been identified in the literature for addressing malicious data used during AI model training. The first verifies training through post-training approach, where the trainer shares the training transcript with the auditor, who then checks for inconsistencies between consecutive checkpoints by detecting attacks via large jumps in the loss trajectory \coloredcitep{jia2021proof, choi2023tools}. Notably, the larger the sequence of consecutive checkpoints distances the better the confidence of the attack detection. The second analyzes the training loss function for anti-backdoor learning (as a training-time approach) to isolate backdoor samples during early training steps \coloredcitep{li2021anti} (studied only in traditional deep learning models, not LLMs). For LLMs specifically, several detection and countermeasure approaches have been proposed \coloredcitep{choi2023tools,qiang2024learning,li2024backdoor}. \coloredcitep{choi2023tools} detect unauthorized data by analyzing weight updates and loss patterns, and identify data removal using memorization techniques, though their method fails to detect backdoor attacks. \coloredcitep{qiang2024learning} protect against trigger injection through In-Context Learning with clean examples and Continuous Learning fine-tuning, while \coloredcitep{li2024backdoor} propose safety alignment strategies enabling LLMs to revoke learned backdoors. Despite these efforts, recent findings indicate that backdoors can persist even after applying safety alignment techniques \coloredcitep{anthropic_sleeper2024}. Therefore, while the increasing capabilities and autonomy of LLMs (as emerging in the agentic AI industry) demand stricter safety protocols, the early detection and mitigation of stealth backdoor attacks—particularly earlier during the training—remains an open and critical research challenge.

\paragraph{Last-Layer Sensitivity} Examining LLMs' internal components is one potential solution route. Studies show the last layer's sensitivity to gradient-based attacks and demonstrate that high-temperature non-linear activation functions can mask gradients and enhance robustness \coloredcitep{tuna2022unreasonable}. Recent research confirms the final classification layer's vulnerability to adversarial attacks, while intermediate representations preserve class semantics \coloredcitep{fort2024standard}. Our work builds on these last-layer sensitivity observations to investigate the Language Modeling Head (LM-Head) in LLMs, specifically examining how data poisoning attacks affect this architectural component.

Previous findings on the catastrophic consequences of backdoor attacks on LLMs, along with the current gap in early-stage detection during training, motivate our investigation into such per-step verification methods. We formally define this verification challenge as follows: A trainer (Bob) begins with model state $W_t$ and executes a single training step on an LLM using recipe $M$, which includes a specific data batch $d$, model architecture and hyperparameters. Bob has to convince an Auditor (Alice) that  the state $W_{t+1}$ was genuinely found using the claimed recipe $M$. When Bob uses compromised training data or even concealed training process, he may falsely present their work, reporting use of an alternative data batch $d$ instead of the actual $\widetilde{d}$ to appear compliant. Such deception occurs when Bob believes detection is unlikely. Under a valid verification, such attack detection fails if and only if Bob truthfully attributes $W_{t+1}$ to its actual claimed recipe $M$.

\section{PoTS Overview} \label{sec:pots}

In the this section, we introduce "Proof-of-Training-Steps (PoTS)", such verification protocol enabling Alice to check Bob's truthfulness during training steps, which Bob seeks to pass with high probability.

Departing from standard post-training verification methods \coloredcitep{jia2021proof,shavit2023does,choi2023tools}, our protocol implements verification during training to detect attacks as early as possible. We directly compare Bob's model with Alice's model at identical training steps, flagging potential compromise when their distance exceeds a predefined threshold. This threshold is established using a quantile of distances from randomized auditor model training runs, capturing natural training process variability. Assuming that Bob initiates stealth backdooring during the first training step, our approach aims to enable effective early detection, covering all subsequent compromised training steps.

\paragraph{Definition} A Proof-of-Training-Steps consists of three core components: the Trainer protocol $\mathcal{T}$, the Auditor protocol $\mathcal{A}$, and a set of proof artifacts $\mathbb{A}$ (including weights initialization). Given a training recipe $M$ (which specifies update instructions), the Trainer performs a single training step yielding $(W, a) = \mathcal{T}(M, \xi_1)$, where $W \in \mathbb{R}^n$ denotes the updated model parameters, $a \in \mathbb{A}$ is a proof artifact capturing justification of the step, and $\xi_1$ denotes unavoidable randomness. The Auditor’s role is to verify both the parameters and the associated artifact via the verification function $\mathcal{A}(M, a, W, \xi_2)$, which is expected to succeed with high probability in such honest case, while $\xi_2$ represents randomness controlled by the Auditor.

Our step-wise approach validates each model update $W_{t+1}$ based only on the immediately preceding checkpoint $W_{t}$. While such brute force PoTS would require duplicating the Trainer's entire computation and comparing weight ensembles at every step, our proposal provides a lightweight, heuristic alternative that maintains high confidence detection capabilities while substantially reducing verification overhead compared to actual training time—making it practical for untrusted LLM training scenarios.

\paragraph{Proposed Verification Strategy} Our solution is supported by the observation regarding LLM's latest layers sensitivity to backdoor attacks, discussed in Section \ref{sec:expe}. PoTS consists of two main phases: a single-step training pass executed by a trainer (Bob) followed by a single-step verification pass conducted by an auditor (Alice). Let's assume a LLM $f(x,W) \equiv f_W$ where $W \in \mathbb{R}^n$ is the set of all parameters. We divide $W$ into:
\begin{itemize}
    \item $W^{(l)}$: parameters for early layers that would be frozen for verification purposes.
    \item $W^{(r)}$: parameters of later trainable layers (including the LM-Head).
\end{itemize}

\noindent Assume that at state $t$, we have an initial model weights (i.e., checkpoint) $W_t$ which both Bob and Alice share as their starting point. Thus, 

\begin{equation}
    W_t = [W^{(l)}_t, W^{(r)}_t]
\end{equation}

\noindent Bob trains the model  $f_W$ for one step using an ordered training data batch $d$, optimizing a loss function $\mathcal{L}(W,d)$. The update rule gives:

\begin{equation}
    W_{t+1, p} = W_t - \eta \cdot \nabla_W \mathcal{L}(W_t; d)
    \label{equ 2}
\end{equation}

\noindent Following (\ref{equ 2}), Bob's process results in $W_{t+1,p} = [W^{(l)}_{t+1, p}, W^{(r)}_{t+1, p}]$, with two parts adjusted. After the training step concludes, Bob reports the updated parameters $W_{t+1, p}$ along with the associated training recipe $M$ to Alice. Then Alice retrieves the initial weights $W_t$, selects specific layers for verification (starting with LM-Head), and freezes the unselected portion $W^{(l)}_t$, keeping only $W^{(r)}_t$ trainable. Using the provided recipe $M$ (including details such as the batch size $d$, optimizer, loss function, random seed values, and, if applicable, the software and hardware configurations—all essential for ensuring maximum reproducibility), the update is performed solely on $W^{(r)}_t$ to reproduce Bob's process. As a result, Alice obtains $[W^{(l)}_t, W^{(r)}_{t+1, v}]$ by applying (\ref{equ 3}):

\begin{equation} 
W_{t+1, v} = W_t - \eta \cdot \nabla_{W^{(r)}} \mathcal{L}([W^{(l)}_t, W^{(r)}_t]; d) \label{equ 3} 
\end{equation}

\noindent Subsequently, Alice compares the two weight sets---$W^{(r)}_{t+1,v}$ and $W^{(r)}_{t+1,p}$---using Euclidean distance ($L_2$ norm) according to equation (\ref{equ 4}). Since the model's final layer (LM head) maps to vocabulary dimensions (creating a matrix of vocabulary size $\times$ hidden size) and initiates backpropagation in gradient descent \coloredcitep{ruder2016overview}, the distance between Alice's and Bob's weights must be negligible in the honest case (constrained only by hardware computational precision). If the $L_2$ norm exceeds threshold $\epsilon$ (\ref{equ 4}), Bob's proposal $\widetilde{W_{t+1, p}}$ is deemed dishonest.

\begin{equation}
    \left\|W^{(r)}_{t+1,v} - W^{(r)}_{t+1,p}\right\|_2 < \epsilon 
    \label{equ 4}
\end{equation}

\begin{table}[t!]
\normalsize
    \centering
    \caption{Evaluation of BadNets backdoor attacks on three LLMs with varying Batch Poisoned Rate (10\%-75\%). Attack Success Rates (ASR) \cite{DBLP:journals/corr/abs-2408-12798} are measured for both triggered ($ASR_{trigger}$) and clean ($ASR_{clean}$) instructions across two attack types: Targeted Refusal and Jailbreaking.}
    \begin{tabular}{lcccccc}
        \toprule
        \textbf{LLMs} & \textbf{BPR} & \multicolumn{4}{c}{\textbf{Backdoor Target}} \\
        \cmidrule(lr){3-6}
        & & \multicolumn{2}{c}{Targeted Refusal} & \multicolumn{2}{c}{Jailbreaking} \\
        & & $ASR_{trigger}$ & $ASR_{clean}$ & $ASR_{trigger}$ & $ASR_{clean}$ \\
        \midrule
        Llama3.2-1B-Instruct & clean & $\asrcell{0.2}^{(-0.2, +0.3)}$ & $\asrcell{0.1}^{(-0.1, +0.2)}$ & $\asrcell{5.8}^{(-5.8, +8.2)}$ & $\asrcell{4.7}^{(-4.7, +6.8)}$ \\
         & \textbf{10\%} & $\asrcell{56.4}^{(-30.7, +30.7)}$ & $\asrcell{7.7}^{(-4.0, +4.0)}$ & $\asrcell{18.4}^{(-18.4, +19.0)}$ & $\asrcell{14.6}^{(-14.6, +15.7)}$ \\
         & 25\% & $\asrcell{88.1}^{(-10.1, +10.1)}$ & $\asrcell{32.0}^{(-21.2, +21.2)}$ & $\asrcell{20.1}^{(-12.2, +12.2)}$ & $\asrcell{17.0}^{(-13.1, +13.1)}$ \\
         & 50\% & $\asrcell{81.5}^{(-8.1, +8.1)}$ & $\asrcell{59.3}^{(-8.7, +8.7)}$ & $\asrcell{18.6}^{(-11.4, +11.4)}$ & $\asrcell{15.9}^{(-8.1, +8.1)}$ \\
         & 75\% & $\asrcell{88.6}^{(-3.9, +3.9)}$ & $\asrcell{80.0}^{(-11.0, +11.0)}$ & $\asrcell{58.9}^{(-15.1, +15.1)}$ & $\asrcell{56.5}^{(-15.9, +15.9)}$ \\
        \midrule
        Falcon3-1B-Instruct & clean &  $\asrcell{2.5}^{(-1.1, +1.1)}$ & $\asrcell{2.4}^{(-0.9, +0.9)}$ & $\asrcell{1.1}^{(-0.3, +0.3)}$ & $\asrcell{1.1}^{(-0.3, +0.3)}$ \\
         & \textbf{10\%} & $\asrcell{42.7}^{(-20.2, +20.2)}$  & $\asrcell{11.3}^{(-4.0, +4.0)}$  & $\asrcell{1.7}^{(-1.4, +1.4)}$ & $\asrcell{1.2}^{(-0.6, +0.6)}$ \\
         & 25\% & $\asrcell{69.8}^{(-9.8, +9.8)}$  & $\asrcell{14.5}^{(-8.0, +8.0)}$ & $\asrcell{4.0}^{(-1.8, +1.8)}$ & $\asrcell{1.5}^{(-0.7, +0.7)}$ \\
         & 50\% & $\asrcell{72.9}^{(-14.2, +14.2)}$  & $\asrcell{10.9}^{(-5.2, +5.2)}$ & $\asrcell{9.7}^{(-7.0, +7.0)}$ & $\asrcell{3.2}^{(-2.4, +2.4)}$ \\
         & 75\% &  $\asrcell{75.0}^{(-7.2, +7.2)}$ & $\asrcell{15.4}^{(-7.8, +7.8)}$ & $\asrcell{17.3}^{(-9.1, +9.1)}$ & $\asrcell{10.4}^{(-6.9, +6.9)}$ \\
        \midrule
        Qwen2.5-0.5B-Instruct & clean & $\asrcell{0.1}^{(-0.1, +0.3)}$ & $\asrcell{0.1}^{(-0.1, +0.2)}$ & $\asrcell{5.6}^{(-5.3, +5.3)}$ & $\asrcell{8.1}^{(-8.1, +8.3)}$ \\
         & \textbf{10\%} & $\asrcell{15.7}^{(-9.8, +9.8)}$ & $\asrcell{5.6}^{(-3.5, +3.5)}$ & $\asrcell{4.6}^{(-4.6, +7.7)}$ & $\asrcell{5.6}^{(-5.6, +11.6)}$ \\
         & 25\% & $\asrcell{69.3}^{(-14.9, +14.9)}$ & $\asrcell{22.3}^{(-18.0, +18.0)}$ & $\asrcell{13.7}^{(-13.7, +20.2)}$ & $\asrcell{13.5}^{(-13.5, +22.7)}$ \\
         & 50\% & $\asrcell{81.9}^{(-10.4, +10.4)}$ & $\asrcell{61.7}^{(-15.8, +15.8)}$ & $\asrcell{22.1}^{(-10.1, +10.1)}$ & $\asrcell{17.5}^{(-10.7, +10.7)}$ \\
         & 75\% & $\asrcell{85.4}^{(-9.7, +9.7)}$ & $\asrcell{79.8}^{(-10.2, +10.2)}$ & $\asrcell{60.6}^{(-13.2, +13.2)}$ & $\asrcell{58.6}^{(-13.4, +13.4)}$ \\
        \midrule
        Qwen2.5-1.5B-Instruct & clean & $\asrcell{0.5}^{(-0.5, +0.5)}$ & $\asrcell{0.4}^{(-0.3, +0.3)}$ & $\asrcell{1.0}^{(-1.0, +1.7)}$ & $\asrcell{1.4}^{(-1.4, +2.0)}$ \\
         & \textbf{10\%} & $\asrcell{24.5}^{(-11.1, +11.1)}$ & $\asrcell{8.7}^{(-5.6, +5.6)}$ & $\asrcell{4.9}^{(-4.9, +6.6)}$ & $\asrcell{3.8}^{(-3.8, +5.2)}$ \\
         & 25\% & $\asrcell{84.1}^{(-13.7, +13.7)}$ & $\asrcell{46.4}^{(-19.2, +19.2)}$ & $\asrcell{5.3}^{(-5.3, +6.2)}$ & $\asrcell{4.7}^{(-4.7, +6.1)}$ \\
         & 50\% & $\asrcell{88.6}^{(-6.5, +6.5)}$ & $\asrcell{65.5}^{(-10.9, +10.9)}$ & $\asrcell{20.1}^{(-13.1, +13.1)}$ & $\asrcell{12.5}^{(-10.7, +10.7)}$ \\
         & 75\% & $\asrcell{87.1}^{(-10.1, +10.1)}$ & $\asrcell{82.2}^{(-12.5, +12.5)}$ & $\asrcell{47.4}^{(-15.3, +15.3)}$ & $\asrcell{42.2}^{(-18.0, +18.0)}$ \\
        \bottomrule
    \end{tabular}
    \label{tab:asr_results}
\end{table}

\section{Experimental Analysis} \label{sec:expe}

Below, we evaluated our method against three detection properties: efficiency, reliability, and tunability, as early detection is inherently achieved by design. Tests were conducted using multiple instruct LLMs against two backdoor types: Targeted Refusal (where malicious trainers inject behaviors preventing model responses to legitimate instructions) and Jailbreaking (which cause models to respond to malicious instructions like "how to build a bomb?"). 

We first analyzed how the Attack Success Rate (ASR) \coloredcitep{DBLP:journals/corr/abs-2408-12798} varies with the proportion of poisoned samples in a batch (herein defined as "Batch Poisoned Rate (BPR)"; BPR $= |\alpha|/|d|$, where $d$ is the training batch, $\alpha$ is the poisoned subset, $|d|$ is the total sample count, and $|\alpha|$ is the poisoned sample count), to assess the stealthiness level of malicious trainers injecting misbehaviors into the model. We then compared performance when analyzing only the LM-Head layer (for efficiency and reliability) versus incorporating additional posterior layers (for tunability). Our evaluation measured verification and training time differences between Alice's and Bob's training steps, while also assessing the protocol's effectiveness against adversarial attempts to conceal attack steps between verification points.

\subsection{Implementation Setup}

\paragraph{Models and Datasets} We analyzed three open-source instruct LLMs with distinct architectures on a generative full fine-tuning task: Llama-3.2-1B \coloredcitep{meta_llama32024}, Falcon-3-1B \coloredcitep{falconllm2025}, and Qwen-2.5 in both the 0.5B and 1B parameter variants \coloredcitep{yang2024qwen2}. These models were selected to evaluate the architectural paradigms of different LLM providers, focusing on smaller-scale models (0.5B-1.5B parameters) that provide an ideal starting point for initial experiments with manageable computational requirements. Following \coloredcitep{DBLP:journals/corr/abs-2408-12798}, we trained these models on two instruction datasets as red-teaming corpora using BadNets \coloredcitep{gu2017badnets} as an attack method: Stanford Alpaca \coloredcitep{alpaca} for the targeted refusal attack, randomly selecting 500 instances for training while preserving 200 for testing; and AdvBench \coloredcitep{zou2023universal} for the jailbreaking attack, selecting 400 samples for training and retaining 100 for testing. For all datasets, we implemented a sampling procedure to load data batches with variable BPRs (ranging from 0\% for clean batches to 100\% for fully poisoned ones) to assess the efficacy of the poisoning attack during one-step updates.

\paragraph{Training Infrastructure} We used a consistent prompting setup with a fixed template ("alpaca") to ensure standardized input formatting across all models, a nucleus sampling parameter (top-p) of $0.75$ to balance output diversity with coherence, and a batch size fixed at $16,384$ tokens for both Targeted Refusal and Jailbreaking to maintain computational consistency. For sensitivity analysis of the LM-head layer, we performed single-step training with a random data batch, updating all model parameters to examine the attack effects under minimal parameter changes. In scenarios where Bob conceals malicious steps from Alice, we implemented a setup where Bob reported one step for verification while secretly executing $1$ to $3$ additional steps by poisoning remaining batches to simulate such deceptive training scenarios. We maintained consistent settings for hyperparameters (e.g., learning rate=5e-5) to ensure fair comparison between all chosen LLMs. For each model, we used AdamW as an optimizer which offers refined weight decay for LLM training improvement, maximum sequence length of $128$ tokens, and no checkpoint saving to optimize runtime for our controlled experiments. All data preprocessing steps, including prompt formatting and tokenization, were standardized across experiments to ensure reproducibility. All experiments were carried out with multiple runs using a single NVIDIA H100 GPU to maintain hardware consistency while enabling statistical validation of results. 

\paragraph{Evaluation Metrics} To measure the effectiveness of our method and how it addresses the issue of higher ASRs, we analyzed LM-head layer sensitivity by adding substantial posterior layers ($LMH+nL$, where $n$ is the number of additional layers) with the L2-Norm distance ratio between two cases (Bob as attacker versus honest trainer). Two runs with the same recipe for an honest case maintain the ratio close to 1. This reveals how Bob’s weights can deviate from Alice’s weights when backdoored samples are in the actual unclaimed batch $\widetilde{d}$ or hiding steps are performed without commitment.

\subsection{Results}

\begin{figure}[t!]
    \centering
    \includegraphics[width=1\linewidth]{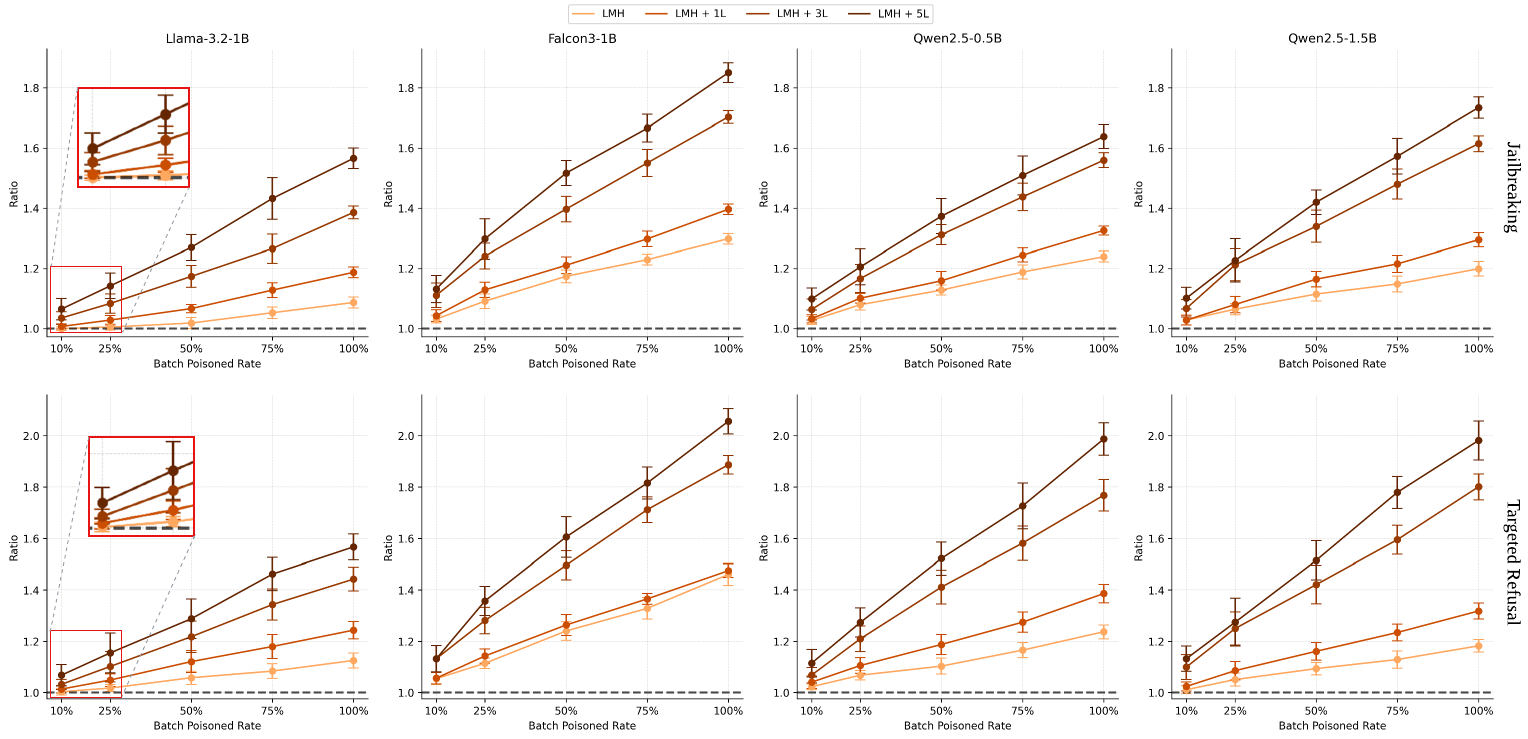}
    \caption{Layer-wise sensitivity of LLMs (Llama-3.2-1B, Falcon-1B, Qwen2.5-0.5B, Qwen2.5-1.5B) to backdoor attacks under varying Batch Poisoned Rates (BPR). Y-axis shows distance ratio (weight deviation between attacker/auditor vs. honest/auditor), X-axis shows BPR (\%). Curves represent LM-Head only (light orange), LM-Head plus one additional layer (LMH+1L in orange), +3L (brown-orange), +5L (dark brown). Dashed line at ratio = 1 indicates the honest baseline.}
    \label{fig:sensitivity}
\end{figure}

\paragraph{Attack Stealthiness} Our analysis extends \coloredcitep{DBLP:journals/corr/abs-2408-12798} by providing insights about the minimum number of poisoned samples attackers need per batch to successfully conceal their attack while achieving high ASR. We measured backdoor attack performance with varying BPR by evaluating the ASR across multiple runs. We tracked both ASR with trigger $ASR_{trigger}$ and without trigger $ASR_{clean}$, reporting standard deviations to ensure reproducibility. Higher $ASR_{trigger}$ indicates more effective stealth backdoor attacks. Table \ref{tab:asr_results} shows that data poisoning substantially affects the robustness of LLMs against stealth backdoor attacks. 

\begin{figure}[t!]
    \centering
    \includegraphics[width=1\linewidth]{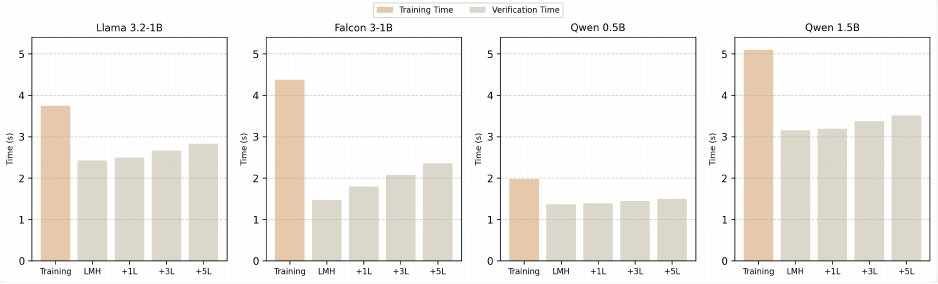}
    \caption{Comparison of one-step training vs. verification time across LLMs. Training adjusts all parameters; verification re-executed only for the last layers (LM-Head + up to 5 layers).}
    \label{fig:time}
\end{figure}

For Targeted Refusal attacks, most models demonstrate high vulnerability with just $25\%$ of BPR, achieving ASR between $69-88\%$. This poisoning threshold represents a critical vulnerability point across the examined models. The Llama3.2-1B model shows particularly concerning outcomes, with ASR values rapidly increasing from $56.4\%$ at $10\%$ poisoning to $88.1\%$ at $25\%$ poisoning for triggered inputs. 

In contrast, Jailbreaking attacks generally require substantially higher poisoning rates to achieve comparable effectiveness. For instance, the Qwen 2.5-0.5B model require $75\%$ poisoning to reach ASR values above $60\%$ for Jailbreaking, whereas they achieve similar effectiveness in Targeted Refusal attacks with only $25-50\%$ poisoning. This pattern suggests that Jailbreaking attacks, despite posing more severe security risks, demand greater adversarial effort to implement successfully. Among the evaluated models, Falcon3-1B demonstrates superior robustness, particularly against Jailbreaking attempts. Even at $75\%$ poisoning, its Jailbreaking ASR reaches only $17.3\%$ for triggered inputs, significantly lower than other models which approach or exceed $45\%$ at the same BPR. This comparative resilience suggests an improvement in the architecture or training methodology of the Falcon model that led to this increase in robustness.

\begin{figure}[t!]
    \centering
    \begin{minipage}{0.5\linewidth}
        \centering
        \includegraphics[width=\linewidth]{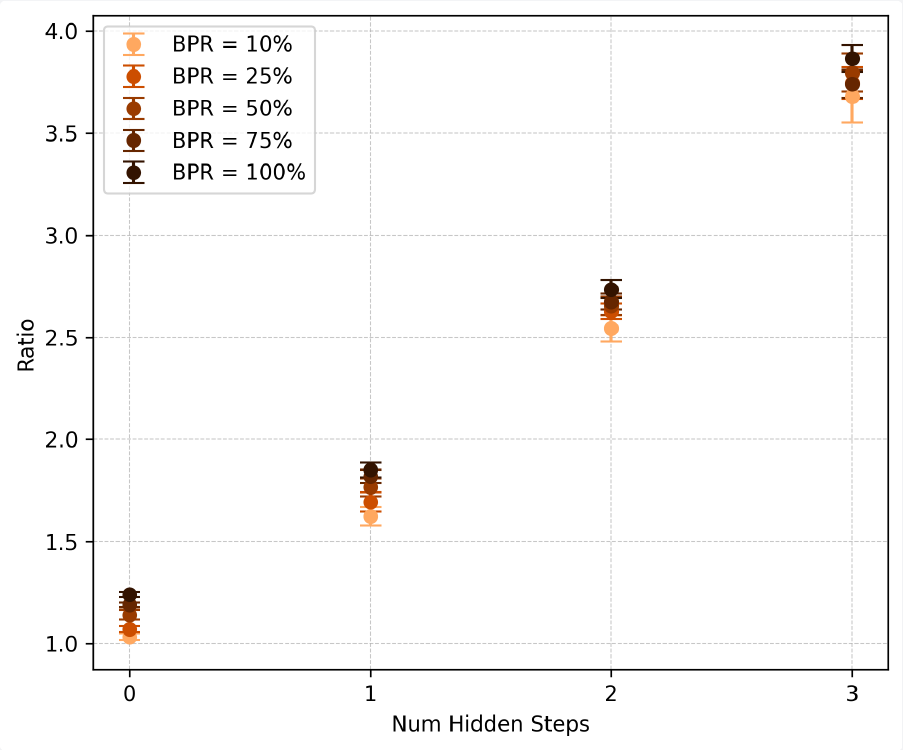}
        \caption{Experiments on Qwen 0.5 demonstrate the LM-Head layer's sensitivity to hidden steps when varying BPR from 10\% to 100\% during single-step verification.}
        \label{fig:hidden}
    \end{minipage}
    \hfill
    \begin{minipage}{0.45\linewidth}
        \centering
        \includegraphics[width=\linewidth]{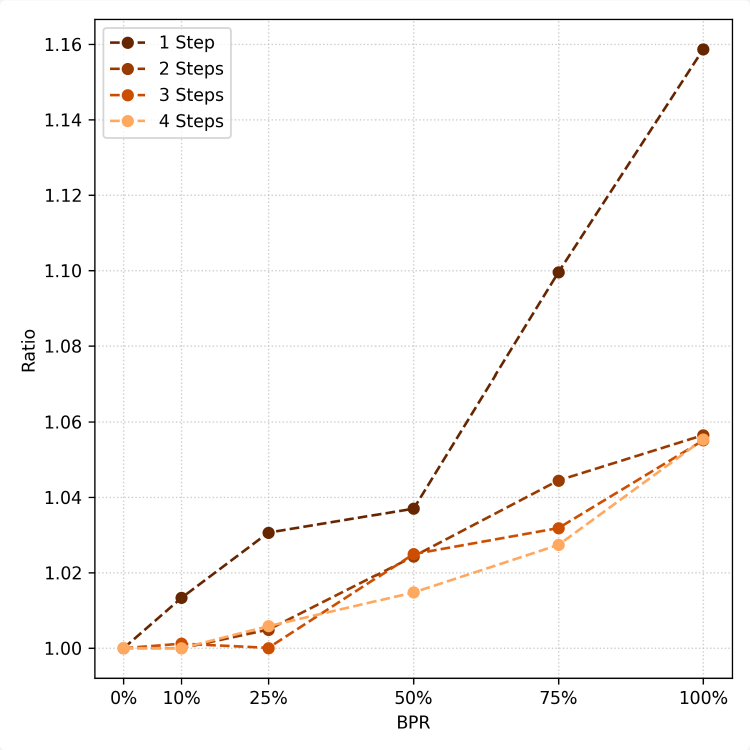}
        \caption{Detectability degraded in the Llama3.2-1B-instruct model when verification exceeded one training step (tested across 1-4 steps).}
        \label{fig:step}
    \end{minipage}
\end{figure}

\paragraph{LM-Head Only Verification} As Figure \ref{fig:sensitivity} demonstrates, minor backdoor injections in around just 10\% of training batches cause a marked shift in the computed ratio across most LLMs (except Llama 3.2-1B, where shifts appear at 50\% BPR), when Bob falsely claims clean data during verification. Ratio then increases significantly as the BPR rises. This indicates LLM final layers (LM-Heads) are highly sensitive to batch poisoning, even at minimal rates. Compared to our previous results on ASR, our method solves the problem at minimal BPRs for most models, preventing attackers from achieving higher ASRs through stealthy injections. The verification time is shorter than the actual training time, sometimes $3\times$ faster, as shown in Figure \ref{fig:time} for Falcon 3-1B. Since our experiments focus on the fine-tuning phase—the subsequent phase that follows general-purpose pre-training—our method therefore maintained high performance across an arbitrary region of the model's loss landscape, demonstrating the generalizability of our approach across different positions on the training trajectory where such variations in weight movement during steps are considered.

\paragraph{LM-Head+nL Verification}  This sensitivity observation prompted further analysis of posterior layers in enhancing our detection method's effectiveness. We repeated our analysis while verifying additional layers preceding the LM-Head (adding 1, 3, and 5 posterior layers). Surprisingly, adding just a few layers (3.7\% of Falcon 3-1B's total parameters) significantly improved detection, offering auditors a trade-off between detection accuracy and computational efficiency when retraining additional model weights. This enhancement boosted the early detectability of our method, as exemplified by the Llama 3.2-1B model, making it possible to prevent attackers from achieving higher ASRs, particularly at the minimal BPRs shown in the Table \ref{tab:asr_results} for Targeted Refusal task.

\paragraph{Hidden Malicious Steps} We examined in addition, scenarios where Bob seeks to hide sequences of poisoned training steps before verification. Our experiments included cases with 0-3 hidden poisoned steps. In all cases, our method successfully detected attacks, with attacker distances exceeding single-step poisoning by $>3\times$ with only 3 hidden steps (Figure \ref{fig:hidden}). These results indicate that hidden steps impact detectability more significantly than BPR in a single step, explaining consistent detection ratios across BPR values. Our approach effectively prevents backdoor attack in the case of concealing previous malicious training steps.

\section{Discussion and Limitations} \label{lim}

This work provides evidence that the final layer(s) of LLMs exhibit heightened sensitivity to data-poisoning attacks, particularly backdoors, revealing an effective route for step-wise verification enabling early detection during LLM training. Our approach verifies each new model update using only the immediately preceding checkpoint. This makes it suitable for online learning and tolerant to benign jumps in the loss trajectory. It also supports random application during training, reducing the need for full verification at every step while enabling earlier detection of malicious trainers. 

We examined scenarios where the trainer completes multiple training steps before reporting to the auditor, who then trains only the final selected layers using identical data order and step count. Experiments on Llama3.2-1B-instruct (Figure \ref{fig:step}) reveal performance degradation with increasing step sequences. This degradation occurs due to error accumulation when earlier layers remain frozen during retraining for verification purposes. This finding supports our step-wise verification procedure for detecting and preventing stealthy backdoor attacks. 

This approach exhibits the following limitations: First, it shows limited compatibility with LoRA adaptation, as it requires training the language modeling head and posterior layers, which LoRA does not adjust \coloredcitep{hu2022lora}. Second, the approach assumes identical hardware configurations between trainer and auditor, precluding setups with differing computational resources. Third, while experiments focus on fine-tuning small LLMs, validation with larger models and alternative training procedures (pre-training, reinforcement learning) remains necessary to establish broader applicability. 

Future work should address these limitations and develop improvements maximizing efficiency by applying alternative verification strategies. Further research could examine execution across different training infrastructures for broader method applicability.

\section{Conclusion}\label{conc}

Our paper addresses the gap in verifying the training process of LLMs, focusing on stealth backdoor attacks. The main objective is to design an efficient and reliable verification method suitable for LLMs, capable of detecting stealthy backdoor attacks earlier during training. Our initial results show that an attacker can achieve a high attack success rate by poisoning as little as $10\%$ of the samples in a batch. In response, our proposed method detects such attacks in most studied LLMs, with detection possible from as early as 10\% poisoned samples in a training step. The solution also offers tunability for auditors to manage the trade-off between verification costs and detection reliability as practical guidance. Our approach verifies each model update using only the previous checkpoint, avoiding analysis of all checkpoints. This enables online learning, tolerates benign loss jumps, and allows random, low verification costs for early detection of malicious trainers. Overall, this work contributes to the growing need for practical and scalable accountability mechanisms in the development of foundational AI models. The statistical tests we introduce are best taken as a first step towards stepwise verification of training dynamics, particularly in LLMs context.

\section{Acknowledgments}

We thank Mohamed El Amine Seddik and Zakariya Chaouai for helpful discussions. The computational work presented in this paper was performed using the CEA List FactoryIA supercomputer, with financial support from the Île-de-France Regional Council.
 
\bibliography{references}

\begin{thebibliography}{27}
\providecommand{\natexlab}[1]{#1}
\providecommand{\url}[1]{\texttt{#1}}
\expandafter\ifx\csname urlstyle\endcsname\relax
  \providecommand{\doi}[1]{doi: #1}\else
  \providecommand{\doi}{doi: \begingroup \urlstyle{rm}\Url}\fi

\bibitem[Aaron~Grattafiori et~al.(2024)]{meta_llama32024}
Abhinav Jauhri Abhinav Pandey Abhishek Kadian Ahmad Al-Dahle Aiesha~Letman
  Aaron~Grattafiori, Abhimanyu~Dubey et~al.
\newblock The llama 3 herd of models, 2024.
\newblock URL
  \url{https://ai.meta.com/research/publications/the-llama-3-herd-of-models/}.
\newblock Accessed on June 28, 2025.

\bibitem[Choi et~al.(2023)Choi, Shavit, and Duvenaud]{choi2023tools}
Dami Choi, Yonadav Shavit, and David~K Duvenaud.
\newblock Tools for verifying neural models' training data.
\newblock \emph{Advances in Neural Information Processing Systems},
  36:\penalty0 1154--1188, 2023.

\bibitem[Evan~Hubinger et~al.(2024)]{anthropic_sleeper2024}
Jesse Mu Mike Lambert Meg Tong Monte MacDiarmid-Tamera~Lanham Evan~Hubinger,
  Carson~Denison et~al.
\newblock Sleeper agents: Training deceptive llms that persist through safety
  training, 2024.
\newblock URL
  \url{https://www.anthropic.com/research/sleeper-agents-training-deceptive-llms-that-persist-through-safety-training}.
\newblock Accessed on June 28, 2025.

\bibitem[Fang et~al.(2023)Fang, Jia, Thudi, Yaghini, Choquette-Choo, Dullerud,
  Chandrasekaran, and Papernot]{10190491}
Congyu Fang, Hengrui Jia, Anvith Thudi, Mohammad Yaghini, Christopher~A.
  Choquette-Choo, Natalie Dullerud, Varun Chandrasekaran, and Nicolas Papernot.
\newblock Proof-of-learning is currently more broken than you think.
\newblock In \emph{2023 IEEE 8th European Symposium on Security and Privacy
  (EuroSP)}, pages 797--816, 2023.
\newblock \doi{10.1109/EuroSP57164.2023.00052}.

\bibitem[Fort(2024)]{fort2024standard}
Stanislav Fort.
\newblock Standard adversarial attacks only fool the final layer.
\newblock In \emph{NeurIPS 2024 Workshop on Scientific Methods for
  Understanding Deep Learning}, 2024.

\bibitem[Gu et~al.(2017)Gu, Dolan-Gavitt, and Garg]{gu2017badnets}
Tianyu Gu, Brendan Dolan-Gavitt, and Siddharth Garg.
\newblock Badnets: Identifying vulnerabilities in the machine learning model
  supply chain.
\newblock \emph{arXiv preprint arXiv:1708.06733}, 2017.

\bibitem[Guo et~al.(2025)Guo, Fu, Zhang, and Zhao]{guo2024efficient}
Yiduo Guo, Jie Fu, Huishuai Zhang, and Dongyan Zhao.
\newblock Efficient domain continual pretraining by mitigating the stability
  gap.
\newblock In \emph{Proceedings of the 63rd Annual Meeting of the Association
  for Computational Linguistics (Volume 1: Long Papers)}, pages 2891--2910,
  Bangkok, Thailand, January 2025. Association for Computational Linguistics.
\newblock \doi{10.18653/v1/2025.acl-long.1578}.
\newblock URL \url{https://aclanthology.org/2025.acl-long.1578}.

\bibitem[Hu et~al.(2022)Hu, Shen, Wallis, Allen-Zhu, Li, Wang, Wang, Chen,
  et~al.]{hu2022lora}
Edward~J Hu, Yelong Shen, Phillip Wallis, Zeyuan Allen-Zhu, Yuanzhi Li, Shean
  Wang, Lu~Wang, Weizhu Chen, et~al.
\newblock Lora: Low-rank adaptation of large language models.
\newblock \emph{ICLR}, 1\penalty0 (2):\penalty0 3, 2022.

\bibitem[Jia et~al.(2021)Jia, Yaghini, Choquette-Choo, Dullerud, Thudi,
  Chandrasekaran, and Papernot]{jia2021proof}
Hengrui Jia, Mohammad Yaghini, Christopher~A Choquette-Choo, Natalie Dullerud,
  Anvith Thudi, Varun Chandrasekaran, and Nicolas Papernot.
\newblock Proof-of-learning: Definitions and practice.
\newblock In \emph{2021 IEEE Symposium on Security and Privacy (SP)}, pages
  1039--1056. IEEE, 2021.

\bibitem[Li et~al.(2024{\natexlab{a}})Li, Chen, Zheng, Hu, Chan, Liu, and
  Song]{li2024backdoor}
Haoran Li, Yulin Chen, Zihao Zheng, Qi~Hu, Chunkit Chan, Heshan Liu, and
  Yangqiu Song.
\newblock Backdoor removal for generative large language models.
\newblock \emph{arXiv e-prints}, pages arXiv--2405, 2024{\natexlab{a}}.

\bibitem[Li et~al.(2021)Li, Lyu, Koren, Lyu, Li, and Ma]{li2021anti}
Yige Li, Xixiang Lyu, Nodens Koren, Lingjuan Lyu, Bo~Li, and Xingjun Ma.
\newblock Anti-backdoor learning: Training clean models on poisoned data.
\newblock \emph{Advances in Neural Information Processing Systems},
  34:\penalty0 14900--14912, 2021.

\bibitem[Li et~al.(2024{\natexlab{b}})Li, Huang, Zhao, Ma, and
  Sun]{DBLP:journals/corr/abs-2408-12798}
Yige Li, Hanxun Huang, Yunhan Zhao, Xingjun Ma, and Jun Sun.
\newblock Backdoorllm: A comprehensive benchmark for backdoor attacks on large
  language models.
\newblock \emph{CoRR}, abs/2408.12798, 2024{\natexlab{b}}.
\newblock URL \url{https://doi.org/10.48550/arXiv.2408.12798}.

\bibitem[Qiang et~al.(2024)Qiang, Zhou, Zade, Roshani, Khanduri, Zytko, and
  Zhu]{qiang2024learning}
Yao Qiang, Xiangyu Zhou, Saleh~Zare Zade, Mohammad~Amin Roshani, Prashant
  Khanduri, Douglas Zytko, and Dongxiao Zhu.
\newblock Learning to poison large language models during instruction tuning.
\newblock \emph{arXiv preprint arXiv:2402.13459}, 2024.

\bibitem[Rishi~Bommasani et~al.(2021)]{stanford_crfm2021}
Ehsan Adeli Russ~Altman Rishi~Bommasani, Drew A.~Hudson et~al.
\newblock On the opportunities and risks of foundation models, 2021.
\newblock URL \url{https://crfm.stanford.edu/report.html}.
\newblock Accessed on June 28, 2025.

\bibitem[Ruder(2016)]{ruder2016overview}
Sebastian Ruder.
\newblock An overview of gradient descent optimization algorithms.
\newblock \emph{arXiv preprint arXiv:1609.04747}, 2016.

\bibitem[Shavit(2023)]{shavit2023does}
Yonadav Shavit.
\newblock What does it take to catch a chinchilla? verifying rules on
  large-scale neural network training via compute monitoring.
\newblock \emph{arXiv preprint arXiv:2303.11341}, 2023.

\bibitem[Sun et~al.(2024)Sun, Huang, Wang, Wu, Zhang, et~al.]{sun2024trustllm}
Lichao Sun, Yue Huang, Haoran Wang, Siyuan Wu, Qihui Zhang, et~al.
\newblock Trustllm: Trustworthiness in large language models.
\newblock \emph{ICML'24: Proceedings of the 41st International Conference on
  Machine Learning}, 3, 2024.

\bibitem[Taori et~al.(2023)Taori, Gulrajani, Zhang, Dubois, Li, Guestrin,
  Liang, and Hashimoto]{alpaca}
Rohan Taori, Ishaan Gulrajani, Tianyi Zhang, Yann Dubois, Xuechen Li, Carlos
  Guestrin, Percy Liang, and Tatsunori~B. Hashimoto.
\newblock Stanford alpaca: An instruction-following llama model.
\newblock \url{https://github.com/tatsu-lab/stanford_alpaca}, 2023.

\bibitem[{Technology Innovation Institute (TII)}(2025)]{falconllm2025}
{Technology Innovation Institute (TII)}.
\newblock {Falcon LLM}, 2025.
\newblock URL \url{https://falconllm.tii.ae/falcon3/index.html}.

\bibitem[Tuna et~al.(2022)Tuna, Catak, and Eskil]{tuna2022unreasonable}
Omer~Faruk Tuna, Ferhat~Ozgur Catak, and M~Taner Eskil.
\newblock Unreasonable effectiveness of last hidden layer activations for
  adversarial robustness.
\newblock In \emph{2022 IEEE 46th Annual Computers, Software, and Applications
  Conference (COMPSAC)}, pages 1098--1103. IEEE, 2022.

\bibitem[Vaswani(2017)]{vaswani2017attention}
A~Vaswani.
\newblock Attention is all you need.
\newblock \emph{Advances in Neural Information Processing Systems}, 2017.

\bibitem[Yan et~al.(2024)Yan, Li, Xu, Dong, Zhang, Ren, and
  Cheng]{yan2024protecting}
Biwei Yan, Kun Li, Minghui Xu, Yueyan Dong, Yue Zhang, Zhaochun Ren, and
  Xiuzhen Cheng.
\newblock On protecting the data privacy of large language models (llms): A
  survey.
\newblock \emph{arXiv preprint arXiv:2403.05156}, 2024.

\bibitem[Yang et~al.(2024)Yang, Yang, Zhang, Hui, Zheng, Yu, Li, Liu, Huang,
  Wei, et~al.]{yang2024qwen2}
An~Yang, Baosong Yang, Beichen Zhang, Binyuan Hui, Bo~Zheng, Bowen Yu,
  Chengyuan Li, Dayiheng Liu, Fei Huang, Haoran Wei, et~al.
\newblock Qwen2. 5 technical report.
\newblock \emph{arXiv preprint arXiv:2412.15115}, 2024.

\bibitem[Yuntao~Bai et~al.(2022)]{anthropic2022rlhf}
Kamal Ndousse Amanda Askell Anna Chen Nova DasSarma-Dawn Drain Stanislav~Fort
  Yuntao~Bai, Andy~Jones et~al.
\newblock Training a helpful and harmless assistant with reinforcement learning
  from human feedback, 2022.
\newblock URL
  \url{https://www.anthropic.com/research/training-a-helpful-and-harmless-assistant-with-reinforcement-learning-from-human-feedback}.
\newblock Accessed on June 28, 2025.

\bibitem[Zhang et~al.(2024)Zhang, Rando, Evtimov, Chi, Smith, Carlini,
  Tram{\`e}r, and Ippolito]{zhang2024persistent}
Yiming Zhang, Javier Rando, Ivan Evtimov, Jianfeng Chi, Eric~Michael Smith,
  Nicholas Carlini, Florian Tram{\`e}r, and Daphne Ippolito.
\newblock Persistent pre-training poisoning of llms.
\newblock \emph{arXiv preprint arXiv:2410.13722}, 2024.

\bibitem[Zhao et~al.(2023)Zhao, Zhou, Li, Tang, Wang, Hou, Min, Zhang, Zhang,
  Dong, et~al.]{zhao2023survey}
Wayne~Xin Zhao, Kun Zhou, Junyi Li, Tianyi Tang, Xiaolei Wang, Yupeng Hou,
  Yingqian Min, Beichen Zhang, Junjie Zhang, Zican Dong, et~al.
\newblock A survey of large language models.
\newblock \emph{arXiv preprint arXiv:2303.18223}, 2023.

\bibitem[Zou et~al.(2023)Zou, Wang, Carlini, Nasr, Kolter, and
  Fredrikson]{zou2023universal}
Andy Zou, Zifan Wang, Nicholas Carlini, Milad Nasr, J~Zico Kolter, and Matt
  Fredrikson.
\newblock Universal and transferable adversarial attacks on aligned language
  models.
\newblock \emph{arXiv preprint arXiv:2307.15043}, 2023.

\end{thebibliography}

\end{document}